\documentclass[aps,prc,reprint,superscriptaddress,nofootinbib]{revtex4-2}

\usepackage{amsmath}
\usepackage{amssymb}
\usepackage{graphicx}
\usepackage{bm}
\usepackage{braket}
\usepackage{mathrsfs}
\usepackage{accents}
\usepackage{longtable}

\begin{document}


\title{
$\alpha$-cluster structures above double shell closures via double-folding potentials from chiral effective field theory
}

\author{Dong Bai}
\email{dbai@tongji.edu.cn}
\affiliation{School of Physics Science and Engineering, Tongji University, Shanghai 200092, China}%

\author{Zhongzhou Ren}
\email[Corresponding author: ]{zren@tongji.edu.cn}
\affiliation{School of Physics Science and Engineering, Tongji University, Shanghai 200092, China}%
\affiliation{Key Laboratory of Advanced Micro-Structure Materials, Ministry of Education, Shanghai 200092, China}


\begin{abstract}

$\alpha$-cluster structures above double shell closures are among the cornerstones for nuclear $\alpha$-cluster physics. 
Semi-microscopic cluster models (SMCMs) are important theoretical models to study their properties.
A crucial ingredient of SMCM is the effective potential between the alpha cluster and the doubly magic nucleus. 
We derive
new double-folding potentials between $\alpha$ clusters and doubly magic nuclei
from soft local chiral nucleon-nucleon potentials given by chiral effective field theory ($\chi$EFT) at the next-to-next-to-leading order.
The $\alpha$-cluster structures in ${}^{8}\text{Be}$, ${}^{20}\text{Ne}$, ${}^{44,52}\text{Ti}$, and ${}^{212}\text{Po}$ are explored to validate these new double-folding potentials.
The $\alpha$ decay of ${}^{104}\text{Te}$ is also studied in the light of recent experimental results.
Our study shows that double-folding potentials from $\chi$EFT are the new reliable effective potentials for the SMCM approach to $\alpha$-cluster structures above double shell closures, with both conceptual and phenomenological merits.

\end{abstract}

\maketitle


\section{Introduction}

$\alpha$-cluster structures in nuclei with two valence protons and two valence neutrons outside double shell closures are among the cornerstones for nuclear $\alpha$-cluster physics.
The simplest example is ${}^{8}\text{Be}$, which has been studied by different methods. 
For instance, it is studied by a new hybrid microscopic model based on nonlocalized clustering and the calculable $R$-matrix theory in Ref.~\cite{Bai:2020hmz},
which provides an exact treatment to the asymptotic boundary conditions in the $\alpha+\alpha$ system.
The next example is ${}^{20}\text{Ne}$. 
Horiuchi and Ikeda pointed out in 1968 that its $K=0^+$ ground-state band and $K=0^-$ band just above the $\alpha$ threshold could be viewed as the ``parity doublets'' produced by the $\alpha+{}^{16}\text{O}$ structure \cite{Horiuchi:1968}.
Later, this picture was generalized to ${}^{44,52}\text{Ti}$, where the $\alpha+{}^{40,48}\text{Ca}$ structures are studied \cite{Michel:1986zz,Michel:1988zz,Wada:1988ik,Ohkubo:1988zz,Yamaya:1990zz,Guazzoni:1993qve,Fukada:2009zz,Bailey:2019iov,Ohkubo:2020esx,Goldkuhle:2019egz}.
$\alpha$-cluster structures in ${}^{212}\text{Po}$ also attract much attention. It is widely accepted that the $\alpha+{}^{208}\text{Pb}$ configuration is crucial for explaining the decay properties of ${}^{212}$Po \cite{Lovas:1998,Astier:2009bs,Suzuki:2010zze,Delion:2012zz}. 
Recently, new experimental results have been reported on $\alpha$ decay of ${}^{104}$Te \cite{Auranen:2018usv,Xiao:2019ywj}, making $\alpha$-cluster structures above ${}^{100}$Sn a new frontier \cite{Ibrahim:2019zlr,Bai:2018giu,Souza:2019eee,Yang:2019oze,Mercier:2020yuv,Clark:2020bum,Baran:2019sqs}.

Semi-microscopic cluster models (SMCMs) are important theoretical models to study $\alpha$-cluster structures above double shell closures \cite{Michel:1998},
where these $\alpha$-cluster states are modeled by two-body systems of $\alpha$ clusters and doubly magic nuclei, e.g., ${}^{8}\text{Be}=\alpha+\alpha$, ${}^{20}\text{Ne}=\alpha+{}^{16}\text{O}$, ${}^{44,52}\text{Ti}=\alpha+{}^{40,48}\text{Ca}$, ${}^{104}\text{Te}=\alpha+{}^{100}\text{Sn}$, and ${}^{212}\text{Po}=\alpha+{}^{208}\text{Pb}$. 
The two constituents are bound together via effective potentials, which are often deep and support not only physical states but also spurious states.
These spurious states are closely related to (almost) Pauli forbidden states in microscopic cluster models.
In SMCMs, the spurious states are identified by the Wildermuth conditions via counting the numbers of nodes in the radial wave functions \cite{Wildermuth:1977} (see Section \ref{NR}).
 The Wildermuth conditions reflect some composite features of $\alpha$ clusters and doubly magic nuclei.
 They are adopted in literature 
to study $\alpha$-cluster structures across the nuclide chart.
In principle, 
some more sophisticated microscopic models
could also be used to study $\alpha$-cluster structures above double shell closures, such as antisymmetrized molecular dynamics (AMD) \cite{KanadaEnyo:1995tb,Kimura:2003uf,Kimura:2004ez}, quantum Monte Carlo method (QMC) \cite{Pastore:2014oda}, lattice effective field theory \cite{Elhatisari:2015iga}, configuration interaction method \cite{Kravvaris:2017nyj}, symmetry-adapted no-core shell model \cite{Dreyfuss:2020lss}, etc.
However, due to heavy computational load, their applications are mainly restricted to light nuclei.
On the contrary, SMCMs can be applied across the nuclide chart, which is an important advantage. 
In SMCMs, pure $\alpha$-cluster configurations are assumed from the beginning.
Therefore, they cannot be used to study the emergence of $\alpha$ clusters from nucleon degrees of freedom by themselves.
However, when combined with experimental data,
 SMCMs could also provide important information on $\alpha$-cluster formation.
For example, $\alpha$-formation probabilities could be extracted by computing the ratios between experimental and SMCM $\alpha$-decay widths.
The SMCM results are important complements to microscopic calculations.

Effective potentials between $\alpha$ clusters and doubly magic nuclei are crucial for SMCMs. Several effective potentials have been proposed,
including double-folding potentials \cite{Ohkubo:1995zz,Ohkubo:1988zz,Mohr:2006yg,Ni:2011zza},
phenomenological potentials in special forms \cite{Buck:1995zza,Wang:2013gk,Bai:2018hbe,Souza:2019eee},
and hybrids between double-folding and phenomenological potentials \cite{Ibrahim:2019zlr}.
Compared to phenomenological potentials, double-folding potentials have closer connections to microscopic models and thus are more favored from the theoretical viewpoint.
The Schmid-Wildermuth force \cite{Schmid:1961}, the Hasegawa-Nagata-Yamamoto force \cite{Hasegawa:1972,Yamamoto:1974}, and the density-independent/dependent Michigan-3-Yukawa (M3Y) interactions \cite{Satchler:1979ni,Khoa:1997,Kobos:1982pdw} are often adopted in literature as the nucleon-nucleon interactions to calculate the double-folding potentials.
Despite some phenomenological success, many of these effective nucleon-nucleon interactions were derived more than forty years ago, by which time our understanding of nucleon interactions was limited.
In the past two decades, tremendous progress has been made in understanding nucleon interactions within the framework of chiral effective field theory ($\chi$EFT), which is a low-energy effective field theory of quantum chromodynamics (QCD) and respects the QCD symmetries (e.g., broken chiral symmetry) \cite{Epelbaum:2008ga,Machleidt:2011zz,Hammer:2019poc}.
$\chi$EFT and the resultant chiral potentials, especially the \emph{realistic} chiral potentials at the next-to-next-to-next-to-leading order ($\text{N}^3\text{LO}$) and beyond, have become the standard inputs for the modern \emph{ab initio} nuclear physics.
On the other hand, it is fair to say that SMCMs have not benefited much from the recent developments of $\chi$EFT.
It is tempting to break this isolation.
Such an attempt will broaden the applicable scope of $\chi$EFT.
%
%
The $\chi$EFT-motivated new effective potentials may also improve the phenomenological agreement between theoretical and experimental results and enhance our confidence in theoretical predictions.

In this work, we derive new double-folding potentials for SMCMs, which are inspired by $\chi$EFT and  act as a bridge  between SMCMs and $\chi$EFT. 
Recently, some encouraging results on double-folding potentials from $\chi$EFT (abbreviated as $\text{DF}_{\!\chi}$s in the following) have been reported and benchmarked explicitly in ${}^{16}\text{O}+{}^{16}\text{O}$, ${}^{12}\text{C}+{}^{12}\text{C}$, and ${}^{12}\text{C}+{}^{16}\text{O}$ elastic scatterings and fusion reactions \cite{Durant:2017ibn,Durant:2020qzj}. 
It is widely accepted that $\alpha$ clustering is a surface phenomenon in the low-density regions of finite nuclei \cite{Brink:1973,Girod:2013}.
As a result, the overlap between the $\alpha$ cluster and the core nucleus is not overlarge in $\alpha$-cluster states. 
It is thus reasonable to expect that $\text{DF}_{\!\chi}$s are also applicable to $\alpha$-cluster structures above double shell closures.
In this work, we adopt the natural units $\hbar=c=1$.

\begin{widetext}

\section{Theoretical Framework}
\label{DFP}

In practice, $\chi$EFT gives different realizations of chiral potentials. 
Many of them are nonlocal in the coordinate space and thus are unfriendly to double-folding calculations. 
Exceptionally, Refs.~\cite{Gezerlis:2013ipa,Gezerlis:2014zia} construct the  local chiral potentials consistently up to the next-to-next-to-leading order ($\text{N}^2\text{LO}$) by exploiting the Fierz rearrangement freedom and local regularization schemes.
At the $\text{N}^2\text{LO}$, the local chiral nucleon-nucleon potentials are given by
\begin{align}
&V_\text{chiral}(\bm{r})=V_\text{L}(\bm{r})\left\{1-\exp[-(r/R_0)^4]\right\}+V_\text{S}(\bm{r}),\label{VChiralT}\\
&V_\text{L}(\bm{r})=V_\text{C}(r)+W_\text{C}(r)\bm{\tau}_1\!\cdot\!\bm{\tau}_2+[V_\text{S}(r)+W_\text{S}(r)\bm{\tau}_1\!\cdot\!\bm{\tau}_2]\bm{\sigma}_1\!\cdot\!\bm{\sigma}_2+[V_\text{T}(r)+W_\text{T}(r)\bm{\tau}_1\!\cdot\!\bm{\tau}_2]S_{12},\\
&V_\text{S}(\bm{r})=(C_\text{S}+C_\text{T}\bm{\sigma}_1\!\cdot\!\bm{\sigma}_2)\,\delta_{R_0}(\bm{r})-(C_1+C_2\bm{\tau}_1\!\cdot\!\bm{\tau}_2)
\Delta\delta_{R_0}(\bm{r})-(C_3+C_4\bm{\tau}_1\!\cdot\!\bm{\tau}_2)\,\bm{\sigma}_1\!\cdot\!\bm{\sigma}_2\Delta\delta_{R_0}(\bm{r})\nonumber\\
&\qquad\ \,+\frac{C_5}{2}\frac{\partial_r\delta_{R_0}(\bm{r})}{r}\bm{L}\!\cdot\!\bm{S}+(C_6+C_7\bm{\tau}_1\!\cdot\!\bm{\tau}_2)
\Bigg\{(\bm{\sigma}_1\!\cdot\!\widehat{\bm{r}})(\bm{\sigma}_2\!\cdot\!\widehat{\bm{r}})
\left[\frac{\partial_r\delta_{R_0}(\bm{r})}{r}\!-\!\partial_r^2\delta_{R_0}(\bm{r})\right]-\!\bm{\sigma}_1\!\cdot\!\bm{\sigma}_2\frac{\partial_r\delta_{R_0}(\bm{r})}{r}\Bigg\},\label{VChiralS}
\end{align}
with $\bm{\sigma}_i$ and $\bm{\tau}_i$ being the Pauli matrices in spin and isospin space, $S_{ij}=3(\bm{\sigma}_i\!\cdot\widehat{\bm{r}})(\bm{\sigma}_j\!\cdot\widehat{\bm{r}})-\bm{\sigma}_i\!\cdot\!\bm{\sigma}_j$ being the tensor operator, and $\bm{L}\!\cdot\!\bm{S}$ being the spin-orbit operator. 
$\delta_{R_0}(\bm{r})=\frac{1}{\pi\Gamma(3/4)R_0^3}\exp[-(r/R_0)^4]$ is the regularized delta function, with $R_0$ being the regularization scale in the coordinate space. The expressions for $V_\text{C,S,T}(r)$ and $W_\text{C,S,T}(r)$ are given in Ref.~\cite{Gezerlis:2014zia}. $C_{\text{S},\text{T},1,\cdots,7}$ are the low-energy constants (LECs)
in the contact sector of $\chi$EFT. 
They generally determine the short-range behavior of chiral nucleon-nucleon potentials.
For the proton-proton pairs, the Coulomb potentials should also be included. 

The double-folding potential between the $\alpha$ cluster and the core nucleus is given by \cite{Khoa:1994zz}
\begin{align}
&U_\text{DF}(\bm{R})=U_\text{D}(\bm{R})+U_\text{Ex}(\bm{R}),\label{UDF}\\
&U_\text{D}(\bm{R})=\!\!\sum_{i,j=p,n}\!\int\!\mathrm{d}^3{r}_\alpha\!\!\int\!\mathrm{d}^3{r}_C\,\rho_\alpha^i(\bm{r}_\alpha)V_\text{D}^{ij}(\bm{s})\rho_{C}^{j}(\bm{r}_{C}),\\
&U_\text{Ex}(\bm{R})=\!\!\sum_{i,j=p,n}\!\int\!\mathrm{d}^3r_\alpha\!\!\int\!\mathrm{d}^3r_C\,\rho^i_\alpha(\bm{r}_\alpha,\bm{r}_\alpha+\bm{s})V_\text{Ex}^{ij}(\bm{s})
\rho^{j}_C(\bm{r}_{C},\bm{r}_{C}-\bm{s})\exp(i\bm{k}_\text{rel}\!\cdot\!\bm{s}/A_\text{red}),\label{UEx}
\end{align}
with $\bm{s}=\bm{R}+\bm{r}_{C}-\bm{r}_\alpha$ being the relative coordinate between two nucleons in the $\alpha$ cluster and the core nucleus, $\rho^{p,n}_{\alpha{(C)}}(\bm{r}_{\alpha{(C)}})$ being the proton and neutron density distributions of the $\alpha$ cluster (core nucleus), $m_\text{red}=m_\alpha m_{C}/(m_\alpha+m_{C})$ and $A_\text{red}=m_\text{red}/m_N$ being the reduced mass and the reduced mass number, and $m_N$ being the average nucleon mass. $k_\text{rel}(R)=\sqrt{2A_\text{red}m_N(E_\text{CM}-U_\text{DF}(R))}$ is the relative momentum, with $E_\text{CM}$ being the energy in the center-of-mass (CM) frame. $\rho^{p,n}_{\alpha(C)}(\bm{r}_{\alpha(C)},\bm{r}_{\alpha(C)}\pm\bm{s})$ are the density matrix elements estimated by the realistic localization approximations \cite{Khoa:1994zz}.
%
%
$V_\text{D(Ex)}^{ij}(\bm{s})$ is the nucleon-nucleon interaction in the direct (exchange) channel. For the $\alpha$ + doubly magic nucleus systems, only the central parts of local chiral nucleon-nucleon potentials make contributions. 
Both the $\alpha$ particle and the heavier doubly magic nucleus have saturated spin-isospin configurations,
which
suppress the contributions from spin-orbit and tensor forces.
Thus, we have \cite{Durant:2017ibn}
\begin{align}
&V_\text{D,Ex}^{pp,nn}(s)=\frac{1}{4}\left[V^{01}(s)\pm3V^{11}(s)\right],\\
&V_\text{D,Ex}^{pn,np}(s)=\frac{1}{8}\left[\pm V^{00}(s)+V^{01}(s)+3V^{10}(s)\pm3V^{11}(s)\right],
\end{align}
with $V^{ST}(s)\equiv\braket{SM_STM_T|V_\text{chiral}(s)|SM_STM_T}$.
The local chiral nucleon-nucleon potentials in Eqs.~\eqref{VChiralT}-\eqref{VChiralS} respect Galileon and isospin symmetry.
As a result, $V^{ST}(s)$ does not depend on $M_S$ and $M_T$.

\end{widetext}

\section{Numerical Results}
\label{NR}

\begin{figure}

  \includegraphics[width=\linewidth]{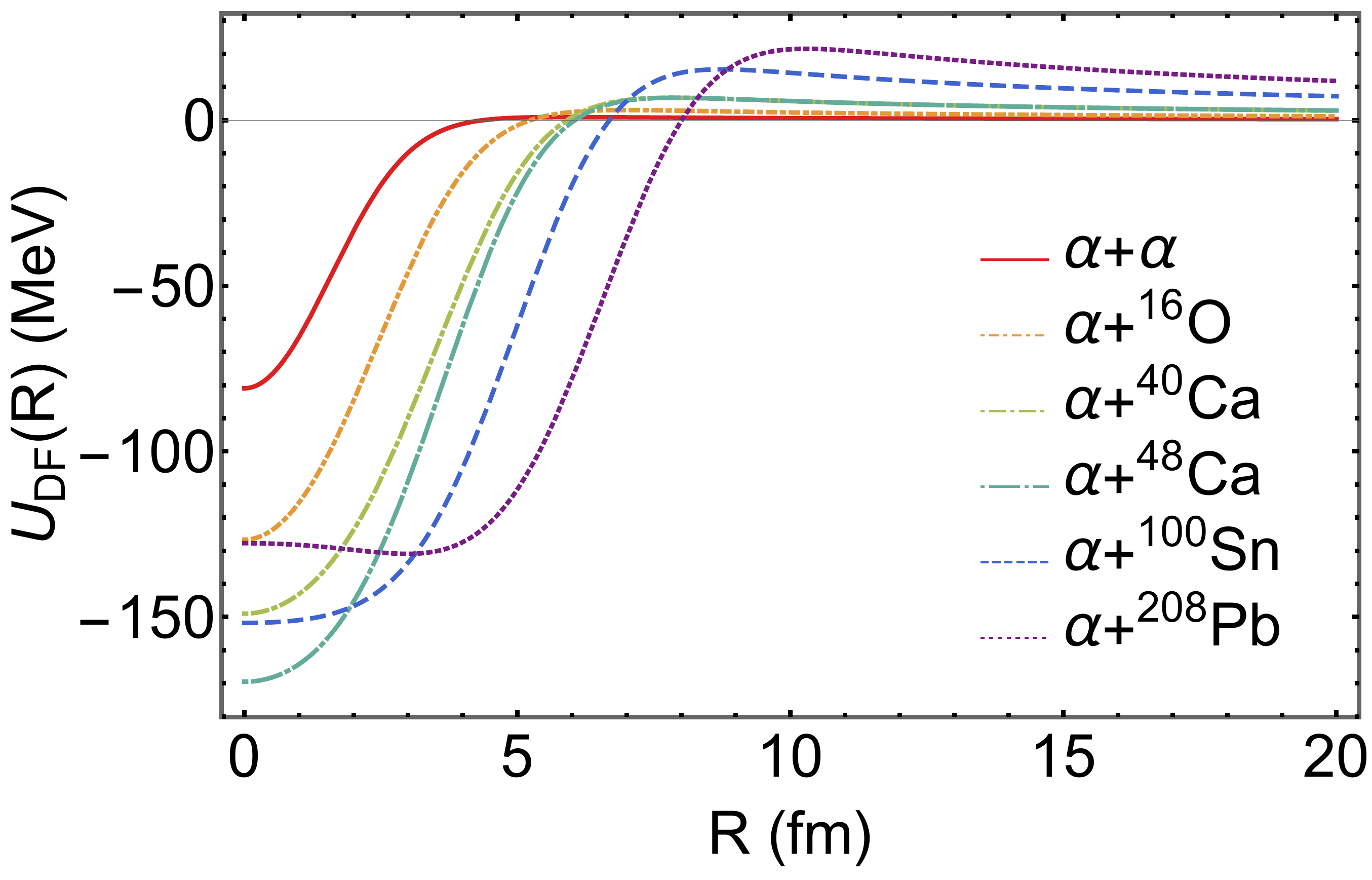}
  
  \caption{The double-folding potentials $U_\text{DF}(R)$ for various $\alpha+\text{doubly magic nucleus}$ systems from soft local chiral nucleon-nucleon potentials at the $\text{N}^2\text{LO}$. See text for details. 
}
  \label{UDFDemo}
  
  \vskip -0.5cm
  
\end{figure}

\begin{table*}

\caption{Theoretical and experimental results for the $0^+_1$, $2^+_1$, and $4^+_1$ states of ${}^{8}$Be. The experimental $\alpha$-decay widths $\Gamma_\alpha^\text{exp}$ are taken from Ref.~\cite{Tilley:2004zz}. $\Gamma_\alpha^{\chi}$ is the $\alpha$-decay width given by SMCM $+$ $\text{DF}_{\!\chi}$.
$R_\text{rel}^\chi\equiv\sqrt{\braket{R^2}}$ is the root-mean-square (RMS) relative distance between two $\alpha$ clusters.
The $0^+_1$ state is studied by using MTPA, while the $2^+_1$ and $4^+_1$ states are studied by using CSM.}
\label{Tab8Be}

\begin{center}

\vskip -0.15cm

\begin{tabular}{cccccccccc}

\hline
\hline
{Nucleus} 			& {$G$} 	& $L^{\pi}$ 	& $\lambda_{NL}$ 	& $\Gamma^\text{exp}_\alpha$ 	& $\Gamma^{\chi}_\alpha$ & $R^\chi_\text{rel}$ \\[-0.25ex]
 & & & & [MeV] & [MeV] & [fm] \\
\hline

${}^{8}\text{Be}$ 	& 4 		& $0^+_1$ 	& 1.4340275 		& $(5.57\pm0.25)\!\times\!10^{-6}$ 	& $6.09\!\times\!10^{-6}$  	 		& 5.33 \\
				& 		& $2^+_1$ 	& 1.402429 		& $1.513\pm0.015$ 				& 1.72 			 				 					& $2.59+0.31i$ \\
				& 		& $4^+_1$ 	& 1.46325 		& $\approx3.5$ 				& 3.18 							 					& $2.90+0.80i$ \\

\hline
\hline

\end{tabular}

\end{center}

\vskip -0.5cm

\end{table*}

\begin{table*}

\caption{Theoretical and experimental results for the $K^\pi=0^+$ and $K^\pi=0^-$ bands of ${}^{20}$Ne. 
$\Gamma_\alpha^\text{exp}$ and $B(\text{E2}\!\downarrow)_\text{exp}$ are the $\alpha$-decay width and reduced electric quadrupole transition strength from Ref.~\cite{Tilley:1998wli}. 
$\Gamma_\alpha^{\chi}$, $B(\text{E2}\!\downarrow)_{\chi}$, $P_\alpha^\chi\equiv\Gamma_\alpha^\text{exp}/\Gamma_\alpha^\chi$, and $R_\text{rel}^\chi$ 
are the theoretical results of the $\alpha$-decay width, the $B(\text{E2}\!\downarrow)$ value, the $\alpha$-formation probability, and the RMS relative distance between the $\alpha$ cluster and ${}^{16}$O given by SMCM $+$ $\text{DF}_{\!\chi}$.
$P_\alpha^\text{AMD}$ is the theoretical $\alpha$-formation probability from AMD \cite{Kimura:2003uf}.
$P_\alpha^\text{h}$ and $B(\text{E2}\!\downarrow)_\text{h}$ are the $\alpha$-formation probability and the $B(\text{E2}\!\downarrow)$ value given by the hybrid potentials in Ref.~\cite{Ibrahim:2019zlr}.
The $K^\pi=0^+$ band and the $1^-_1$, $3^-_2$ states in the $K^\pi=0^-$ band are studied by using MTPA, while the other states are studied by using CSM.}
\label{Tab20Ne}

\begin{center}

\vskip -0.4cm

\hspace*{-1cm}
\begin{tabular}{ccccccccccccc}

\hline
\hline
{Nucleus} & {$G$} & $L^{\pi}$ & $\lambda_{NL}$  & $\Gamma^\text{exp}_\alpha$ & $\Gamma^{\chi}_\alpha$ & $P_\alpha^\text{AMD}$ & ${P}^{\chi}_\alpha$ & $P^\text{h}_\alpha$ & $B(\text{E2}\!\downarrow)_\text{exp}$ & $B(\text{E2}\!\downarrow)_{\chi}$ & $B(\text{E2}\!\downarrow)_\text{h}$ & $R^\chi_\text{rel}$ \\[-0.25ex]
 & & & & [MeV] & [MeV] & & & & [W.u.] & [W.u.] & [W.u.] & [fm] \\
\hline

${}^{20}\text{Ne}$		& 8 		& $0^+_1$ 		& 1.1518273 			& 								& 							& 0.70	& 				&				&  						& 				& 				& 3.79 \\
					& 		& $2^+_1$ 		& 1.1370375 			& 								& 							& 0.68	&				&				& $20.3\pm1.0$   			& 13.0 			& 14.3 			& 3.80 \\
					& 		& $4^+_1$ 		& 1.13190728 			& 								& 							& 0.54	&  				&				& $22\pm2$  				& 17.1 			& 18.5			& 3.73 \\
					& 		& $6^+_1$ 		& 1.1172209 			& $(1.1\pm0.2)\!\times\!10^{-4}$		& $3.64\!\times\!10^{-4}$  			& 0.34  	& $0.30\pm0.05$	& $0.19\pm0.04$  	& $20\pm3$ 				& $15.2$   		& 15.2			& 3.62 \\
					& 		& $8^+_1$ 		& 1.1674842 			& $(3.5\pm1.0)\!\times\!10^{-5}$  		& $1.98\!\times\!10^{-4}$			& 0.28  	& $0.18\pm0.05$ 	& $0.095\pm0.027$	& $9.0\pm1.3$				& 7.0 			& 7.9		 		& 3.21 \\
					& 9 		& $1^-_1$ 		& 1.1919326 			& $(2.8\pm0.3)\!\times\!10^{-5}$  		& $2.72\!\times\!10^{-5}$			& 0.95  	& $1.03\pm0.11$	& $0.82\pm0.09$	& 						& 				&				& 4.78 \\
					& 		& $3^-_2$ 		& 1.2043384 			& $(8.2\pm0.3)\!\times\!10^{-3}$  		& $7.95\!\times\!10^{-3}$			& 0.93  	& $1.03\pm0.04$	& $0.67\pm0.02$ 	& $50\pm8 $ 				& $41.6$			& 77.0			& $4.84$ \\
					& 		& $5^-_3$ 		& 1.2074985 			& $0.145\pm0.40$ 					& 0.114 						& 0.88  	& $1.27\pm0.35$ 	& $0.73\pm0.20$	&  						& $40.0+9.6i$	& 126.9			& $4.52+0.30i$ \\
					& 		& $7^-_3$ 		& 1.202514 			& $0.110\pm0.010$ 					& 0.314						& 0.71  	& $0.35\pm0.03$ 	& $0.20\pm0.02$	& 						& $27.8+10.7i$	& 154.9			& $4.09+0.32i$ \\
					& 		& $9^-_5$ 		& 1.18843 			& $0.225\pm0.040$  					& 0.354						& 0.70  	& $0.64\pm0.11$ 	& $0.38\pm0.07$	& 						& $14.7+6.5i$	& 36.6			& $3.69+0.23i$ \\
\hline
\hline

\end{tabular}
\end{center}

\vskip -0.75cm

\end{table*}

\begin{table}

\caption{Theoretical and experimental results for the observed states in the $K^\pi=0^+$ and $K^\pi=0^-$ bands of ${}^{44}$Ti. 
The experimental $B(\text{E2}\!\!\downarrow)_\text{exp}$s are taken from Ref.~\cite{Chen:2011jjw}.}
\label{Tab44Ti}

\begin{center}

\begin{tabular}{ccccccccccc}

\hline
\hline
{Nucleus} & {$G$} & $L^{\pi}$ & $\lambda_{NL}$ & $B(\text{E2}\!\downarrow)_\chi$ & $B(\text{E2}\!\downarrow)_\text{exp}$ & $R^\chi_\text{rel}$ \\[-0.25ex]
 & & & & [W.u.] & [W.u.] & [fm] \\
\hline

${}^{44}\text{Ti}$ & 12 & $0^+_1$ & 1.1019607 & & & 4.32 \\
& & $2^+_1$ & 1.0915923 & 9.9 & $13\pm4$ & 4.33 \\
& & $4^+_1$ & 1.0853305 & 13.4 & $30\pm5$ & 4.28 \\
& & $6^+_1$ & 1.08476 & 12.7 & $17.0\pm2.4$ & 4.18  \\
& & $8^+_1$ & 1.078867 & 10.5 & & 4.06  \\
& & $10^+_1$ & 1.0990893 & 6.7 & & 3.85  \\
& & $12^+_1$ & 1.1338819 & 3.0 & & 3.63 \\
 & 13 & $1^-_2$ & 1.1232387 & & & 4.86 \\
 & & $3^-_6$ & 1.116978 & 20.0 & & 4.83 \\
 & & $5^-_3$ & 1.1053947 & 22.1 & & 4.78 \\
 & & $7^-_2$ & 1.0972716 & 20.7 & & 4.67 \\
\hline
\hline

\end{tabular}

\end{center}

\vskip -0.6cm

\end{table}

\begin{table}

\caption{Theoretical and experimental results for the $K^\pi=0^+$ band of ${}^{52}$Ti.
The experimental $B(\text{E2}\!\downarrow)_\text{exp}$s are taken from Ref.~\cite{Goldkuhle:2019egz}.
$B(\text{E2}\!\downarrow)_{\text{WS}^2}$s are the theoretical results given by the phenomenological $\text{WS}^2$ potential in Ref.~\cite{Ohkubo:2020esx}.}
\label{Tab52Ti}

\begin{center}

\vskip -0.25cm

\hspace*{-0.35cm}
\begin{tabular}{ccccccccccc}

\hline
\hline
{Nucleus} & $L^{\pi}$ & $\lambda_{NL}$  & $B(\text{E2}\!\downarrow)_\text{exp}$ & $B(\text{E2}\!\downarrow)_\chi$ & $B(\text{E2}\!\downarrow)_{\text{WS}^2}$ & $R^\chi_\text{rel}$ \\[-0.25ex]
 & & & [W.u.] & [W.u.] & [W.u.] & [fm] \\
\hline

 ${}^{52}\text{Ti}$ 		& $0^+_1$ 		& 0.9656424  		& 						& 					& 			& 4.20 \\
 					& $2^+_1$ 		& 0.9564093  		& $7.5^{+0.4}_{-0.3}$ 		&7.1  				& 9.4			& 4.20 \\
 					& $4^+_1$ 		& 0.9497294  		& $9.5^{+1.4}_{-1.1}$  		& 9.6 				& 12.3		& 4.16  \\
 					& $6^+_1$ 		& 0.9550252  		& $8.7^{+0.6}_{-0.5}$ 		& 8.8					& 11.6		& 4.05  \\
 					& $8^+_1$ 		& 0.9584199  		& $0.76\pm0.09$ 			& 7.1 				& 9.5			& 3.93  \\
 					& $10^+_1$ 		& 0.95336506 		&  						& 5.0 				& 6.6			& 3.80  \\
\hline
\hline

\end{tabular}
\end{center}

\vskip -0.75cm

\end{table}

\begin{table*}

\caption{Theoretical and experimental results for the ground-state band of ${}^{212}$Po. 
The experimental $B(\text{E2})_\text{exp}$s and $\alpha$-decay widths of the $0^+_1$ and $18^+_1$ states are taken from Ref.~\cite{Browne:2005kpt}.
The experimental $\alpha$-decay widths of the $6^+_1$ and $8^+_1$ states are taken from Refs.~\cite{Astier:2010sn,Lieder:1978}.
$P_\alpha^\text{LA}=-1.16066 + 0.20035R_\text{rel}^\chi$ is the $\alpha$-formation probability given by the linear approximation in Refs.~\cite{Bai:2018bjl,Bai:2019jmv}.
$\Gamma_\alpha^{\chi,\text{refined}}=P_\alpha^\text{LA}\Gamma_\alpha^\chi$ is the refined theoretical $\alpha$-decay width.
$\Gamma_\alpha^\text{CDM3Y6}$ is the $\alpha$-decay width given by the double-folding potential from the CDM3Y6 effective nucleon-nucleon interaction with a constant $\alpha$-formation probability $P^\text{CDM3Y6}_\alpha=0.3$ and the renormalization factors $\lambda_{NL}^\text{CDM3Y6}\sim0.5$ \cite{Ni:2011zza}.
}
\label{Tab212Po}

\begin{center}

\vskip -0.3cm

\hspace*{-0.5cm}
\begin{tabular}{ccccccccccccc}

\hline
\hline
{Nucleus} 				& {$G$} 		& $L^{\pi}$ 			& $\lambda_{NL}$ 		& $\lambda_{NL}^\text{CDM3Y6}$	& $P_\alpha^\text{LA}$  & $\Gamma^\text{exp}_\alpha$ & $\Gamma^\chi_\alpha$  			& $\Gamma^{\chi,\text{refined}}_\alpha$		& $\Gamma^{\text{CDM3Y6}}_\alpha$ 		& $B(\text{E2}\!\downarrow)_\text{exp}$ &	 $B(\text{E2}\!\downarrow)_\chi$  & $R_\text{rel}^\chi$ \\[-0.25ex]
 					& 			& 					& 					& 				&     & [MeV] 							& [MeV] 								& [MeV]					& [MeV]				   & [W.u.] 			& [W.u.] 			& [fm] \\
\hline
${}^{212}\text{Po}$ 		& 22 			& $0^+_1$ 			& 1.0459895 			& 0.576				& 0.094			& $(1.53\pm0.01)\!\times\!10^{-15}$ 		& $1.62\!\times\!10^{-14}$ 				&  $1.53\!\times\!10^{-15}$ 		& $2.85\!\times\!10^{-15}$		& 				& 				& 6.26 \\
					& 			& $2^+_1$ 			& 1.0376702 			& 0.572				& 0.099 			&								& $4.18\!\times\!10^{-13}$ 				& $4.12\!\times\!10^{-14}$ 		& $7.31\!\times\!10^{-14}$		&				& 6.3 		 	& 6.29 \\
					& 			& $4^+_1$ 			& 1.031552 			& 0.569				& 0.095			&								& $8.13\!\times\!10^{-13}$ 				& $7.72\!\times\!10^{-14}$ 		& $1.43\!\times\!10^{-13}$	 	&				& 8.8 			& 6.27 \\
					& 			& $6^+_1$ 			& 1.02607 			& 0.565				& 0.085 			& $\left(1.8_{-0.5}^{+1.2}\right)\!\times\!10^{-14}$ 	& $3.19\!\times\!10^{-13}$ 		& $2.72\!\times\!10^{-14}$ 		& $5.52\!\times\!10^{-14}$		& $3.9\pm1.1$ 		& 9.1 			& 6.22  \\
					& 			& $8^+_1$ 			& 1.020155 			& 0.562				& 0.072 			& $\left(1.9_{-0.3}^{+0.4}\right)\!\times\!10^{-15}$ 	& $3.97\!\times\!10^{-14}$ 		& $2.85\!\times\!10^{-15}$ 		& $6.81\!\times\!10^{-15}$		& $2.30\pm0.09$	& 8.7 			& 6.15 \\
					& 			& $10^+_1$ 			& 1.010274 			& 0.556				& 0.059 			&								& $7.28\!\times\!10^{-15}$ 				& $4.30\!\times\!10^{-16}$ 		& $1.23\!\times\!10^{-15}$		& $2.2\pm0.6$		& 7.9 			& 6.09 \\
					& 			& $12^+_1$ 			& 0.993857 			& 0.547				& 0.048 			&								& $5.57\!\times\!10^{-15}$ 				& $2.69\!\times\!10^{-16}$ 		& $9.39\!\times\!10^{-16}$		&				& 7.1 			& 6.03 \\
					& 			& $14^+_1$ 			& 0.982049 			& 0.540				& 0.033 			&								& $1.34\!\times\!10^{-16}$ 				& $4.40\!\times\!10^{-18}$ 		& $2.21\!\times\!10^{-17}$		&				& 5.8 			& 5.96 \\
					& 			& $18^+_1$ 			& 0.95507 			& 0.524				& 0.0034 			& $\left(1.01^{+0.02}_{-0.01}\right)\!\times\!10^{-23}$	& $3.01\!\times\!10^{-21}$ 		& $1.01\!\times\!10^{-23}$ 		& $4.60\!\times\!10^{-22}$		& 				& 				& 5.81 \\[0.2ex]
\hline
\hline

\end{tabular}
\end{center}

\vskip -0.5cm

\end{table*}

In numerical calculations, we take the regularization scale $R_0=1.6\text{ fm}$. The corresponding LECs are given in Ref.~\cite{Durant:2017ibn}.
They are determined by fitting the Nijmegen neutron-proton phase shifts in the ${}^{1}\text{S}_0$, ${}^{3}\text{S}_1$, ${}^{1}\text{P}_1$, ${}^{3}\text{P}_1$, ${}^{3}\text{P}_2$, and ${}^{3}\text{S}_1$-${}^{3}\text{D}_1$ channels at 1, 5, 10, 25, 50, 100, and 150 MeV.
The resultant local chiral nucleon-nucleon potentials are known as soft local chiral nucleon-nucleon potentials, as they have soft cores at the short distance, crucial for the double-folding calculations.
LECs at smaller regularization scales may not be suitable for our current purposes, as they generally give much stronger repulsive cores and may break the mean-field picture behind the double-folding calculations.
We would like to stress that the soft local chiral nucleon-nucleon potentials do not mean to be the realistic nucleon-nucleon potentials fitting the world nucleon scattering data up to 290 MeV at the level of $\chi^2/\text{datum}\approx1$.
One has to go to the next-to-next-to-next-to-leading order $(\text{N}^3\text{LO})$ and beyond to meet that requirement.
On the contrary, the soft local chiral nucleon-nucleon potentials are formulated up to the $\text{N}^2\text{LO}$ only and are aimed at fitting selected nucleon scattering data at low energies.
The proton density distributions for $\alpha$ particle, ${}^{16}$O, ${}^{40,48}$Ca, and ${}^{208}$Pb are taken to be the realistic sums of Gaussians determined by the elastic electron scattering experiments \cite{DeJager:1987qc}, while the neutron density distributions are assumed to be proportional to the proton density distributions for simplicity.
For ${}^{100}$Sn, no elastic electron scattering data are available, and we take the S$\tilde{\text{a}}$o Paulo distributions 
$\rho^{p,n}(r)\!=\!{\rho_0^{p,n}}\!\Big/\!\left[{1+\exp\left(\frac{r-R_{p,n}}{a_{p,n}}\right)}\right]$,
with $R_p=1.81Z^{1/3}-1.12\ \text{fm}$, $R_n=1.49N^{1/3}-0.79\ \text{fm}$, 
$a_p=0.47-0.00083Z\ \text{fm}$, $a_n=0.47+0.00046N\ \text{fm}$ \cite{Chamon:2002mx}.
 $\rho_0^{p,n}$ are determined by $\int\!\mathrm{d}^3r\rho^{p,n}(r)=Z,N$.
The charge radius of ${}^{100}$Sn is found to be 4.58 fm, in good agreement with $4.525$-$4.707$ fm found by \emph{ab initio} self-consistent Green’s function theory \cite{Arthuis:2020toz}.
$E_\text{CM}$ in Eq.~\eqref{UEx} is taken to be $E_\text{CM}=Z_\alpha Z_C/(A_\alpha^{1/3}+A_C^{1/3})$ MeV,
which provides a convenient estimation of the height of the Coulomb barrier between the $\alpha$ cluster and the core nucleus
$\sim{Z_\alpha Z_{C}\,e^2}/{(R_\alpha+R_C)}=Z_\alpha Z_C\,e^2/[r_0(A_\alpha^{1/3}+A_C^{1/3})]\sim{Z_\alpha Z_C}/(A_\alpha^{1/3}+A_C^{1/3})\ \text{MeV}$,
with $r_0=1.31\ \text{fm}$ \cite{Glendenning:2004}.
This is widely used in nuclear reaction studies. 
In the last step of the derivation, we use the approximation $e^2/r_0\approx1$ MeV.


$\text{DF}_{\!\chi}$s are shown for several $\alpha+\text{doubly magic nucleus}$ systems in Fig.~\ref{UDFDemo}. 
The physical properties of $\alpha$-cluster states are obtained by solving the Schr\"odinger equation
$\left[-\frac{\nabla^2}{2m_\text{red}}+U({\bm R})\right]\Psi_{NLM}({\bm R})=E_{NL}\Psi_{NLM}({\bm R})$,
with $U({\bm R})=\lambda_{NL} U_\text{DF,N}({\bm R})+U_\text{DF,C}({\bm R})$.
Here, $U_\text{DF,N}(\bm{R})$ and $U_\text{DF,C}(\bm{R})$ are the nuclear and Coulomb parts of the double-folding potential $U_\text{DF}({\bm R})$. 
$\lambda_{NL}$ is the renormalization factor introduced for phenomenological reasons.
It is determined by reproducing the experimental energy level exactly.
$\text{DF}_{\!\chi}$s are deep potentials in the sense that they support not only physical states but also spurious states.
These spurious states are closely related to the (almost) Pauli forbidden states in microscopic cluster models.
The Pauli forbidden states are null eigenstates of norm operators in resonating group method (RGM) with identical oscillator parameters.
In SMCMs, the spurious states are identified by the Wildermuth conditions \cite{Wildermuth:1977} as $\alpha$-cluster states with $G\equiv2N+L<4$ for ${}^{8}\text{Be}$, $G<8$ for ${}^{20}$Ne, $G<12$ for ${}^{44}$Ti, $G<12$ and $(G,L)=(12,12)$ for ${}^{52}$Ti, $G<16$ and $(G,L)=(16,14), (16,16)$ for ${}^{104}$Te, and $G<22$ and $(G,L)=(22,20), (22,22)$ for ${}^{212}$Po. Here, $N$ is the number of nodes in the $\alpha$-cluster radial wave function (excluding the origin), and $L$ is the orbital angular momentum. We take into account the extra constraints from the occupied proton orbits $0g_{9/2}$, $0h_{11/2}$ in ${}^{100}\text{Sn}$, ${}^{208}\text{Pb}$ and the occupied neutron orbits $0f_{7/2}$, $0g_{9/2}$, $0i_{13/2}$ in ${}^{48}\text{Ca}$, ${}^{100}\text{Sn}$, ${}^{208}\text{Pb}$.
For light nuclei, the Wildermuth conditions could be verified explicitly
by solving the eigenvalue problems of the RGM norm operators.
For example, the eigenvalues of the RGM norm operator have been worked out to be $\mu_G=1-2^{2-G}+3\delta_{G,0}$ for the $\alpha+\alpha$ system \cite{Horiuchi:1977}.
It is straightforward to see that the Pauli forbidden states with $\mu_G=0$ satisfy the condition $G=0, 2<4$, 
which is exactly the same as the Wildermuth condition mentioned before.
The problem becomes complicated for heavy nuclei.
Rigorously speaking, 
 for the applications of the Wildermuth conditions to $\alpha+\text{heavy-core}$ models,
the Pauli-forbidden states are not defined clearly,
as the oscillator parameters of the $\alpha$ cluster and the heavy core nucleus are largely different.
Even if the oscillator parameters of the same size are used,
the eigenvalues of the norm kernel for the Pauli-allowed states could be very small due to a large number of nucleons in the core nucleus.
Compared with light nuclear systems such as $\alpha+\alpha$, heavy nuclear systems also have a much larger configuration space,
and the spin-orbit interactions become important.
The $\alpha$-formation probabilities in heavy nuclei are generally smaller than light nuclei.
Therefore,  
it is less straightforward to see 
whether the Wildermuth conditions could simulate the antisymmetrization in realistic heavy nuclei to good accuracy.
In the case of different oscillator parameters, the first eigenstates of the RGM norm operator with small but nonzero eigenvalues are called the almost Pauli forbidden states,
which, by definition, should become the Pauli forbidden states in the limit of identical oscillator parameters.
It is known for light nuclear systems that the almost Pauli
forbidden states correspond to high-lying states with large energy expectation values (i.e., the Pauli resonances) \cite{Horiuchi:1977,Saito:1977,Baye:1992zz,Saito:1973,Kruglanski:1992nuh}.
As a result, they are weakly coupled to the low-lying shell-model and cluster states due to the large energy difference.
This explains the availability of the Wildermuth condition in the presence of the almost Pauli forbidden states for some light nuclear systems.
Similar results may hold for heavy nuclear systems as well.
It might be helpful and illuminative if the energy expectation values of the almost Pauli forbidden states were worked out for these systems.
At present, such calculations are not available yet.
Nevertheless, it is important to continue to examine the Wildermuth conditions in heavy nuclei.
They provide useful references for future microscopic calculations, as well as theoretical motivations to develop 
better semi-microscopic approximations for antisymmetrization.
%
%
%
%
We encounter both bound and resonant states in calculations. 
For the resonant states with $\alpha$-decay widths $\Gamma_\alpha \geq 0.01$ MeV, we obtain their physical properties by using the complex scaling method (CSM) \cite{Moiseyev:2011}.
Applying bound-state concepts (e.g., the relative distance between the $\alpha$ cluster and the doubly magic nucleus, the electric quadrupole transition, etc) to resonant states leads to
imaginary parts in theoretical results, which are interpreted as theoretical uncertainties by Berggren \cite{Berggren:1996kno}. 
In principle, CSM could also be used to study resonant states with narrow $\alpha$-decay widths $<0.01$ MeV. However, we find that for these states the $\alpha$-decay widths given by CSM become unstable for different complex scaling angles in numerical calculations. Thus, we adopt the modified two-potential approach (MTPA) \cite{Gurvitz:2003ai} to study the long-lived resonant states with $\Gamma_\alpha<0.01$ MeV.
In MTPA, the tunneling potential is divided into the inner and outer parts by a separation radius.
The long-lived resonant states are then approximated by bound states supported by the inner part of the tunneling potential,
from whose wave functions the decay widths of the long-lived resonant states are computed.
The $R$-matrix theory \cite{Lane:1948zh,Thompson:2009,Descouvemont:2010cx,Burke:2011} is also commonly used to study long-lived resonant states.
In Ref.~\cite{Gurvitz:2003ai}, the similarities and differences between MTPA and the $R$-matrix theory are discussed in detail.
We try to do the calculation with the $R$-matrix theory as well. The theoretical results agree well with MTPA.
For example, our $R$-matrix calculation gives the $\alpha$-decay width $\Gamma_\alpha^\chi=6.055\times10^{-6}$ MeV for the ${}^{8}$Be ground state, 
which is almost identical to the MTPA result in Table \ref{Tab8Be} with negligible difference $\sim$ 0.6\% \cite{Bai:2020c}.
At last, we would like to mention that we assume pure $\alpha$-cluster configurations in calculating the theoretical $\alpha$-decay widths unless otherwise mentioned.

 \begin{figure}

  \includegraphics[width=\linewidth]{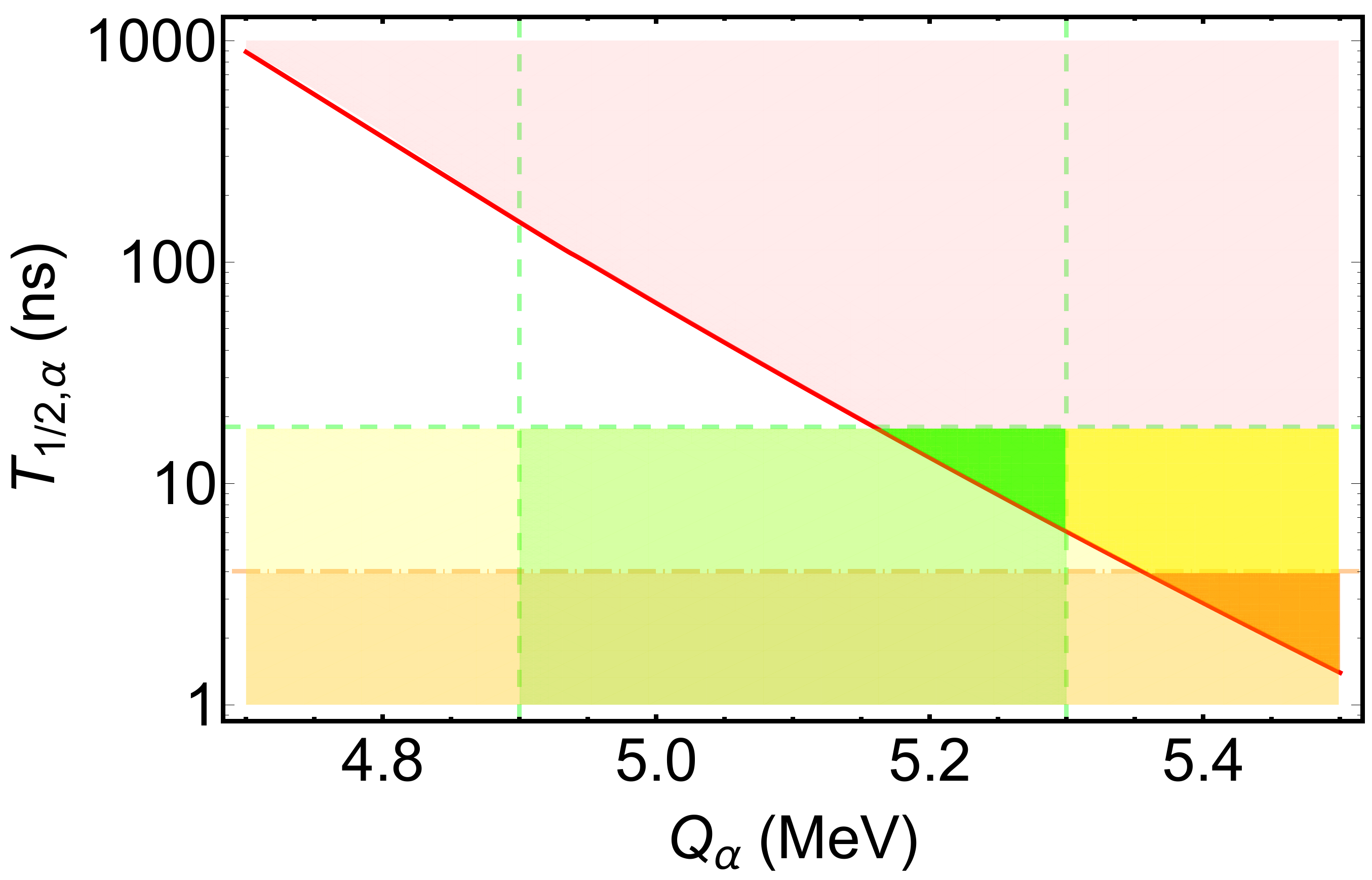}
  
  \caption{
The joint experimental and theoretical analysis on $\alpha$-decay properties of ${}^{104}$Te.
The red solid line shows the relation between the $\alpha$-decay half-life and $Q_\alpha$ given by SMCM $+$ $\text{DF}_{\!\chi}$ with the $\alpha$-formation probability $P_\alpha=1$.
The light green and light yellow regions are the $1\sigma$ and $2\sigma$ bands from Ref.~\cite{Auranen:2018usv}. 
The light orange region is the parameter space allowed by Ref.~\cite{Xiao:2019ywj}.
The light red region estimates the  SMCM $+$ $\text{DF}_{\!\chi}$-allowed parameter space from the condition $P_\alpha\leq1$.
The green, yellow, and orange regions are the overlaps of these four regions.
  }
  \label{104Te}
  
  \vskip -0.5cm
  
\end{figure}

The numerical results for ${}^{8}$Be are given in Table \ref{Tab8Be}. We calculate the $\alpha$-decay widths $\Gamma_\alpha^{\chi}$ and the root-mean-square (RMS) relative distances between two $\alpha$ clusters $R^\chi_\text{rel}\equiv\sqrt{\braket{R^2}}$ for the $0_1^+$, $2_1^+$, $4_1^+$ states. 
The experimental $\alpha$-decay widths $\Gamma_\alpha^\text{exp}$ 
 are listed for comparison. 
The $\alpha$-decay widths given by SMCM $+$ $\text{DF}_{\!\chi}$ are in good agreement with the experimental data, compatible with the dominance of the $\alpha$-cluster configurations in the $0_1^+$, $2_1^+$, and $4_1^+$ states suggested by microscopic models \cite{Bai:2020hmz}.

The numerical results for ${}^{20}$Ne are given in Table \ref{Tab20Ne}. 
We also include the experimental results \cite{Tilley:1998wli}, the AMD results \cite{Kimura:2003uf}, and the phenomenological results from the hybrid potentials \cite{Ibrahim:2019zlr} in the same table for comparison.
The AMD calculations show that the $1^-_1$, $3^-_2$, and $5^-_3$ states have the $\alpha$-formation probabilities $P_\alpha^\text{AMD}\sim 1$.
In other words, these states have almost the pure $\alpha$-cluster structures. 
They can be used as the additional benchmarks besides the $0^+_1$, $2^+_1$, and $4^+_1$ states in ${}^{8}$Be to compare different effective potentials in SMCM without worrying too much about $\alpha$-formation probabilities.
The $\alpha$-decay widths $\Gamma_\alpha^\chi$ given by SMCM $+$ $\text{DF}_{\!\chi}$ agree well with the experimental data for the $1^-_1$, $3^-_2$, and $5^-_3$ states. 
Also, the experimental result of the enhanced $B(\text{E2}\!\downarrow)_\text{exp}$ from the $3^-_2$ state to the $1^-_1$ state is nicely reproduced by our model as shown in Column Eleven.
In Column Eight are the $\alpha$-formation probabilities $P_\alpha^\chi\equiv\Gamma^\text{exp}_\alpha/\Gamma^\chi_\alpha$ extracted by SMCM $+$ $\text{DF}_{\!\chi}$,
which generally agree well with the AMD results.
Here, we estimate the $\alpha$-formation probability as the ratio between the experimental and SMCM $\alpha$-decay widths.
This estimation
 could be understood as follows.
The $\alpha$-formation probability measures the amount of $\alpha$ clustering.
Let's take the $\alpha$-cluster state with $L=0$
as an example.
The discussions could be easily generalized to $\alpha$-cluster states with $L\neq0$.
When $L=0$, the $\alpha$-formation probability is given by 
$P_\alpha\equiv\int\mathrm{d}^3r|G(\bm{r})|^2$
in microscopic models,
with $G(\bm{r})=\braket{\widetilde{\Phi}_{\bm{r}}|\Psi}$ being the $\alpha$-formation amplitude.
${\Psi}$ is the normalized realistic many-body wave function for the target nucleus.
It could be given by a combination of cluster and shell-model components.
$\widetilde{\Phi}_{\bm{r}}=\mathcal{N}^{-1/2}\Phi_{\bm{r}}$ is the normalized wave function for the pure $\alpha$-cluster configuration,
with $\Phi_{\bm{r}}=\mathscr{A}_{Aa}\{\Phi^{(A)}(\bm{\xi}_A)\Phi^{(a)}(\bm{\xi}_a)\delta^{3}(\bm{r}-\bm{r}_{Aa})\}$
and $\mathcal{N}$ being the RGM norm operator
with the kernel $N(\bm{r},\bm{r}')=\braket{\Phi_{\bm{r}}|\Phi_{\bm{r}'}}$.
$\Phi^{(A)}(\bm{\xi}_A)$ and $\Phi^{(a)}(\bm{\xi}_a)$ are the normalized intrinsic wave functions of the core nucleus and the $\alpha$ particle.
The $\alpha$-formation probability satisfies $P_\alpha=\int\mathrm{d}^3r |\braket{\widetilde{\Phi}_{\bm{r}}|\Psi}|^2\leq \!\int\!\mathrm{d}^3r\braket{\widetilde{\Phi}_{\bm{r}}|\widetilde{\Phi}_{\bm{r}}}\braket{\Psi|\Psi}=1$
thanks to the Cauchy inequality.
Let's assume that the experimental $\alpha$-decay width could be reproduced by microscopic models, 
where
the $\alpha$-decay width is written as
$\Gamma_\alpha^\text{exp}=2P_0(Q_\alpha,a)\gamma_0^2(Q_\alpha,a)$,
with $P_0(Q_\alpha,a)=ka/\left|\mathcal{H}^{(+)}_0(\eta,ka)\right|^2$
and $\gamma_0^2(Q_\alpha,a)=2\pi a G(a)^2/m_\text{red}$. 
Here, $a$ is the channel radius in the $R$-matrix calculation,
$k=\sqrt{2m_\text{red}Q_\alpha}$ is the wave number of the $\alpha$ cluster 
in the infinity,
$\eta$ is the Coulomb-Sommerfeld parameter,
and $\mathcal{H}^{(+)}_0(\eta,ka)$ is the $S$-wave outgoing Coulomb-Hankel function.
Although $P_0(Q_\alpha,a)$ and $\gamma_0^2(Q_\alpha,a)$
depend on the channel radius $a$ separately, in principle,
their product $\Gamma_\alpha^\text{exp}$ should not depend on $a$ according to the $R$-matrix theory \cite{Lane:1948zh,Thompson:2009,Descouvemont:2010cx,Burke:2011}.
In practice, due to limited model space and finite numerical precision, 
some residual dependence on the channel radius could appear in numerical calculations.
In that case, it is important to choose the channel radius in an appropriate way.
%
%
%
We would like to stress again that the $\alpha$-formation amplitude $G(\bm{r})$ is normalized to the $\alpha$-formation probability $P_\alpha\leq1$.
On the other hand, 
if it is the pure $\alpha$-cluster configurations that are assumed in theoretical calculations, the $\alpha$-formation amplitudes (aka the $\alpha$-cluster wave functions) are normalized to unity instead of the $\alpha$-formation probability $P_\alpha\leq1$. 
Then, the theoretical $\alpha$-decay width should satisfy $\Gamma_\alpha^\text{th}\approx \Gamma_\alpha^\text{exp}/P_\alpha$.
As mentioned before, SMCMs assume pure $\alpha$-cluster configurations
when calculating the theoretical $\alpha$-decay width $\Gamma_\alpha^\chi$.
Therefore, we can use $P_\alpha^\chi=\Gamma_\alpha^\text{exp}/\Gamma_\alpha^\chi$ to estimate the $\alpha$-formation probability of the target state. 
 In Columns Nine and Twelve, we list the $\alpha$-formation probabilities $P_\alpha^\text{h}$s and the $B(\text{E2}\!\downarrow)$ values $B(\text{E2}\!\downarrow)_\text{h}$s extracted from Ref.~\cite{Ibrahim:2019zlr} based on the hybrid potentials for comparison.
Last but not least, we would like to mention that the RMS relative distances between the $\alpha$ cluster and ${}^{16}$O in Column Thirteen are found to decrease along the $K^\pi=0^+$ and $K^\pi=0^-$ bands in general.
This is also observed in previous studies based on other effective potentials.
It is sometimes referred to as the antistretching effect \cite{Michel:1998}.

The numerical results for $^{44}$Ti are given in Table \ref{Tab44Ti}. 
The experimental data on the $\alpha$-cluster states in ${}^{44}$Ti are limited compared to ${}^{20}$Ne.
Especially, the $\alpha$-decay data are not available from the experimental side, which forbids the extraction of $\alpha$-formation probabilities in SMCM.
Nevertheless, it is found that, similar to ${}^{20}$Ne, $B(\text{E2}\!\downarrow)$s in the $K^\pi=0^-$ band are about two times those in the $K^\pi=0^+$ band,
which might be the hints for prominent $\alpha$-cluster structures in the $K^\pi=0^-$ band.
Moreover, the antistretching effect is also observed in ${}^{44}$Ti.

The numerical results for ${}^{52}$Ti are given in Table \ref{Tab52Ti}. 
The global quantum number is taken to be $G=12$.
Unlike ${}^{44}$Ti, the neutron orbit $0f_{7/2}$ has been occupied by the core nucleus ${}^{48}$Ca. Therefore, the $\alpha$-cluster state with $(G,L)=(12,12)$ is unfavored by the Wildermuth condition,
and the $K^\pi=0^+$ band gets terminated at $L=10$ automatically, which is consistent with experimental data.
$B(\text{E2}\!\downarrow)_\chi$s from SMCM are given in Column Five.
They agree well with the experimental data in Column Four, except for the $8^+_1$ state.
The $8^+_1$ state has its $B(\text{E2}\!\downarrow)_\text{exp}$ be one order of magnitude smaller than $B(\text{E2}\!\downarrow)_\chi$,
indicating that this state is more likely to be a shell-model state. 
In Column Six are $B(\text{E2}\!\downarrow)_{\text{WS}^2}$s from the phenomenological $\text{WS}^2$ potential \cite{Ohkubo:2020esx}.
In comparison, our $\text{DF}_{\!\chi}$s are in better agreement with the experimental data.

The numerical results for ${}^{212}$Po are given in Table \ref{Tab212Po}.
The global quantum number is taken to be $G=22$. 
The proton orbit $0h_{11/2}$ and the neutron orbit $0i_{13/2}$ have been occupied by the core nucleus ${}^{208}$Pb.
As a result, the $\alpha$-cluster states with $(G,L)=(22,20),(22,22)$ are unfavored by the Wildermuth condition,
and the ground-state band gets terminated automatically at $L=18$, consistent with the experimental observation.
The theoretical $\alpha$-decay widths from SMCM with $P_\alpha=1$ are given in Column Eight.
The $\alpha$-formation probability of the $0^+_1$ state is extracted to be $P^\chi_\alpha=\Gamma_\alpha^\text{exp}/\Gamma_\alpha^\chi=0.094$, compatible with $P_\alpha^\text{QWFA}=0.1045$ from quartetting wave function approach \cite{Yang:2019oze}. 
On the other hand, the $\alpha$-formation probability of the $18^+$ state is found to be as tiny as $P^\chi_\alpha=0.0034$, 
suggesting that the shell-model configuration is more important in this state.
Refs.~\cite{Bai:2018bjl,Bai:2019jmv} propose an approximately linear relation between the $\alpha$-formation probability and the RMS relative distance $R_\text{rel}^\chi$ along the band.
Therefore, the $\alpha$-formation probabilities along the ground-state band of ${}^{212}$Po could be estimated by the linear relation $P_\alpha^\text{LA}=-1.16066 + 0.20035R_\text{rel}^\chi$. The refined $\alpha$-decay widths $\Gamma_\alpha^{\chi,\text{refined}}\equiv P_\alpha^\text{LA}\Gamma_\alpha^\chi$ are given in Column Nine, agreeing well with the experimental results. 
The theoretical results given by the CDM3Y6 double-folding potentials from Ref.~\cite{Ni:2011zza} are given in Columns Five and Ten.
In Ref.~\cite{Ni:2011zza}, a constant $\alpha$-formation probability $P^\text{CDM3Y6}_\alpha=0.3$ is used to calculate the theoretical $\alpha$-decay widths.
Such a choice is motivated by a previous microscopic result on the $\alpha$-formation probability in the ground state of ${}^{212}$Po \cite{Varga:1992zz}. 
In this work, we give an improved treatment on the $\alpha$-formation probabilities by taking into consideration the evolution of 
$\alpha$-cluster formation along the ground-state band \cite{Bai:2018bjl,Bai:2019jmv} under the guidance of the latest microscopic result from Ref.~\cite{Yang:2019oze}.
In order to reproduce the experimental energy levels, the renormalization factors $\lambda_{NL}^\text{CDM3Y6}\sim0.5$ are needed in Ref.~\cite{Ni:2011zza},
which deviate sizeably from 1 and deform the potentials in a significant way.
In comparison, in SMCM $+$ $\text{DF}_{\!\chi}$ the renormalization factors $\lambda_{NL}$ deviate only slightly from 1.
The theoretical $\alpha$-decay widths in Table \ref{Tab212Po} are tiny compared to those in Tables \ref{Tab8Be} and \ref{Tab20Ne}.
They are calculated 
by MTPA. 
MTPA has also been used successfully to calculate $\alpha$-decay widths
of other heavy and superheavy nuclei (see, e.g., Ref.~\cite{Qian:2014pbn}),
where the $\alpha$-decay widths can be even smaller than those listed in Table \ref{Tab212Po}.

 
At last, we study the $\alpha$ decay of ${}^{104}$Te.
 Auranen \emph{et al.}~report that  $Q_\alpha({}^{104}\text{Te})=5.1(2)$ MeV,  $T_{1/2,\alpha}({}^{104}\text{Te})<18$ ns \cite{Auranen:2018usv}. 
Later on, Xiao \emph{et al.}~observe two new events compatible with Ref.~\cite{Auranen:2018usv} in an independent experiment and give the constraint $T_{1/2,\alpha}({}^{104}\text{Te})<4$ ns \cite{Xiao:2019ywj}.
But, they cannot fully exclude the possible impacts from $\beta$ decay.
We carry out a joint experimental and theoretical analysis based on these two experimental results and SMCM $+$ $\text{DF}_{\!\chi}$.
The results are given in Fig.~\ref{104Te}.
It is found that most parts of the parameter space allowed by experiments are actually disfavored by SMCM, except the upper right corner. 
If Ref.~\cite{Auranen:2018usv} is considered only,
it is the triangular regions colored in green and yellow (overlapping partially with the green and orange triangular regions) that are favored by the joint analysis with the confidence levels of 68\% and 95\%, respectively.
If both experiments are considered, then
the triangular region colored in orange is most favored by the joint analysis.

\section{Conclusions}
\label{Concl}

In summary, we derive new reliable double-folding potentials for the $\alpha+\text{doubly magic nucleus}$ systems from $\chi$EFT and use them to study the $\alpha$-cluster structures in ${}^{8}$Be, ${}^{20}$Ne, ${}^{44,52}$Ti, and ${}^{212}$Po within the framework of SMCM.
Compared with the existing effective potentials, 
$\text{DF}_{\!\chi}$s have better connections to QCD via $\chi$EFT.
Besides, they give theoretical results in good agreement with experimental data.
$\alpha$-decay properties of ${}^{104}$Te are also studied in the light of two recent experimental results. 
Our study shows that $\text{DF}_{\!\chi}$s are new reliable effective potentials for the SMCM approach to $\alpha$-cluster structures above double shell closures.


\begin{acknowledgments}

We thank Gerd R\"opke and Peter Schuck for useful discussions.
We thank Peter Mohr for helpful communications on ${}^{104}$Te.
Also, we thank the anonymous referee for helpful guidance.
This work is supported by the National Natural Science Foundation of China (Grants No.\ 11947211, No.\ 11905103, No.\ 12035011, No.\ 11535004, No.\ 11975167, No.\ 11761161001, No.\ 11565010, No.\ 11961141003, and No.\ 12022517), by the National Key R\&D Program of China (Contracts No.\ 2018YFA0404403 and No.\ 2016YFE0129300), by the Science and Technology Development Fund of Macau (Grants No.\ 0048/2020/A1 and No.\ 008/2017/AFJ), by the Fundamental Research Funds for the Central Universities (Grant No.\ 22120200101), and by the China Postdoctoral Science Foundation (Grants No.\ 2020T130478 and No.\ 2019M660095).


\end{acknowledgments}

\end{document}